\documentclass[sigconf]{acmart}
\AtBeginDocument{%
  \providecommand\BibTeX{{%
    \normalfont B\kern-0.5em{\scshape i\kern-0.25em b}\kern-0.8em\TeX}}}

\setcopyright{none}

\begin{document}

\title{Private Chat in a Public Space of Metaverse Systems}

\author{Jiarui Chen}
\authornote{Each author contributed equally to this research.}
\affiliation{%
  \institution{National University of Singapore}
  \country{Singapore}
}
\email{e0893429@u.nus.edu}

\author{Xinwei Loo}
\authornotemark[1]
\affiliation{%
  \institution{National University of Singapore}
  \country{Singapore}}
\email{e0550444@u.nus.edu}

\author{Yien Hong}
\authornotemark[1]
\affiliation{%
  \institution{National University of Singapore}
  \country{Singapore}}
\email{ianhong@u.nus.edu}

\author{Anand Bhojan}
\affiliation{%
  \institution{National University of Singapore}
  \country{Singapore} 
}
\email{dcsab@nus.edu.sg}

\renewcommand{\shortauthors}{Bhojan and Chen, et al.}

\begin{abstract}
  With the proliferation of Virtual Reality (VR) technologies and the emergence of the Metaverse, social VR applications have become increasingly prevalent and accessible to the general user base. Serving as a novel form of social media, these platforms give users a unique opportunity to engage in social activities. However, there remains a significant limitation: the inability to engage in private conversations within public social VR environments. Current interactions are predominantly public, making it challenging for users to have confidential side discussions or whispers without disrupting ongoing conversations. To address this gap, we developed Hushhub, a private chat system integrated into the popular social VR platform VRChat. Our system enables users within a shared VR space to initiate private audio conversations selectively, allowing them to maintain awareness and engagement with the broader group discussions. To evaluate the system, we conducted user studies to gather insight and feedback on the efficacy and user experience of the implemented system. The results demonstrate the value and necessity of enabling private conversations within immersive social VR environments, paving the way for richer, more nuanced social interactions.
\end{abstract}

\maketitle

\section{Introduction}

With the rapid advancement and growing popularity of virtual reality (VR) technology, social VR platforms have become significant spaces for digital social interactions, closely resembling real-world gatherings and meetings \cite{maloney2021social, joseph_2021}. These immersive environments allow users to interact through VR headsets, satisfying diverse social needs in increasingly realistic virtual settings \cite{george2021metaverse, sykownik2021most, dzardanova2022virtual}.

Despite their growing adoption, current social VR platforms exhibit a critical limitation: the inability to hold private, parallel conversations akin to whispering or side chats common in real-world group interactions. In face-to-face meetings, participants naturally engage in selective, brief private exchanges without disrupting the broader group discussion, an essential dynamic of real-world social interactions. Such parallel conversations facilitate nuanced exchanges, coordination, and clarification, thereby enriching the overall quality of social interactions \cite{swaab2008pros, sarkar2021promise}. However, this feature is absent in mainstream social VR applications, where all communications remain equally perceivable by all users in the same room or in a world instance \cite{jonas_2019, osborne_2023}.

In social VR settings, the presence of private communication channels can significantly enhance user satisfaction, social presence, and overall engagement by providing additional layers of interaction and context that mirror real-world conversations \cite{barreda2023easily}. Previous studies also highlight the role of self-disclosure in developing relationships with others in the Metaverse and how private channels can encourage users to disclose personal information \cite{hooi2014avatar, sykownik_2022, maloney2020anonymity}. This underscores the importance of creating private communication spaces in VR platforms.

Recognizing the importance of replicating real-world communication dynamics, we developed Hushhub, a system designed to enable selective private conversations within shared VR spaces, an innovation absent in current platforms \cite{jonas_2019}. Unlike current solutions that require isolating users to separate rooms or joining a new world-instance, our implementation mimics natural conversational behavior within a shared environment. Hushhub allows users to seamlessly initiate discreet, private audio channels with nearby users while simultaneously maintaining awareness and engagement with the general public discourse. The design of allowing users in private channels to hear external conversations not only avoid the creation of echo chambers but also satisfy users' need of engaging in meaningful interactions without losing context from existing study \cite{lo2021escape, limbago_2023}.

By granting users greater control over their communication, determining both who can speak to them and whom they wish to address, our system fosters a more convenient environment simulating the real-world dynamics. As such, this research explores the following research questions (RQs):

\begin{enumerate}
    \item  \textbf{RQ1.} What challenges do users face in initiating conversations in public VR spaces, and how does the absence of privacy features impact their ability to communicate effectively? 
    \item  \textbf{RQ2.} How does the ability to conduct private conversations within public VR spaces impact user satisfaction, relationship-building, and overall social experience?
\end{enumerate}

\section{Background and Related Work}

Social virtual reality (VR) platforms have emerged as dynamic environments that support immersive, real-time social interactions. As highlighted by Maloney et al. [\cite{maloney2021social}], these platforms are characterized by spatialized communication, where users engage through proximity-based audio in shared virtual spaces, simulating co-presence and enhancing social presence. While this model fosters a sense of realism and embodiment, it also imposes structural constraints on interaction. 

It is important to differentiate the context of mainstream social VR platforms (such as VRChat or Rec Room) from that of professional or corporate metaverse platforms. These corporate environments, such as Virbela, have indeed implemented mechanisms for private conversations. For example, Virbela features 'Private Volumes,' which are spatially-defined zones that users can enter to conduct audio-isolated meetings, unheard by anyone outside the volume. The use of such platforms for remote work and their impact on communication dynamics has been a subject of recent study \cite{hong2024exploring}. However, the design of these corporate-focused 'private volumes' typically prioritizes audio isolation to prevent distractions, effectively creating a hard barrier between the private group and the public space. This model, while suitable for formal breakout sessions, does not address the need for more fluid, "whisper-like" side conversations common in social gatherings, where users may wish to remain partially aware of the main group's activity. Our work with HushHub is built on this distinction; it is designed not simply to create privacy, but to do so while allowing the user to maintain social awareness of the broader public context (as detailed in Table 1), addressing a specific gap in social VR interactions.

Jonas et al. [\cite{jonas_2019}] conducted a feature analysis of 39 social VR applications. Their findings revealed that no existing application offered user-controlled private voice channels, despite the prevalence of basic features such as muting and blocking. The authors point out that the lack of support for layered or parallel communication — such as private side conversations — limits the range of natural social dynamics that can occur. As a result, current communication systems in social VR are effective for open dialogue but fall short in supporting more complex, context-sensitive interactions found in real-world group settings.

Research in both physical and virtual environments underscores the vital role that private or parallel communication channels play in fostering richer, more nuanced social interactions. Swaab et al. [\cite{swaab2008pros}] conclude that dyadic side conversations can influence group dynamics — enhancing openness to diverse information and improving problem - solving performance in tasks that benefit from faction-forming, while also demonstrating potential drawbacks depending on group norms and task context in real-world face-to-face settings. In digital settings, Sarkar et al. [\cite{sarkar2021promise}] found that the emergence of parallel chat during video meetings can support flexible coordination, inclusivity, and real-time collaboration without disrupting the main dialogue—although it can introduce distraction—illustrating its dual impact on meeting dynamics. Within social VR contexts, preliminary studies explore related dynamics: Maloney et al. [\cite{maloney2020anonymity}] highlight how channels for nuanced communication—such as whispers or private audio—enhance self-disclosure and foster familiarity among users in VR environments. Similarly, Sykownik et al. [\cite{sykownik_2022}] find that allowing personal conversations enables users to share more personal content and navigate social contexts with greater depth—not possible in purely public voice spaces. Together, this body of work reveals how private and concurrent communication channel significantly enriches interaction quality and social presence, both in traditional and VR-mediated settings.

Building on user-centred design insights from prior work, our system follows recommendations for integrating private conversation mechanisms into shared environments. Limbago [\cite{limbago_2023}] conducted formative studies to explore diverse private chat methods—such as "floating icons" and teleport-to-room metaphors—but found that while these options provided privacy, they disrupted immersion and social continuity. Their findings highlighted users’ preference for intuitive, unobtrusive side-conversation methods that maintain presence within the primary virtual space. Recent systems have advanced this direction. The design-space exploration by Limbago et al. [\cite{limbago2025don}] outlined a framework considering multiple dimensions—privacy level, social awareness, modality—emphasizing a fluid transition between public and private dialogue and the importance of awareness cues even while in private chat.

Together, these studies guide our Hushhub design by prioritizing seamless integration—enabling private conversations within ongoing public discourse, minimizing disruption to group dynamics, and preserving social presence—consistently with proven principles for immersive, user-friendly private chat in social VR.

\section{System Design and Implementation}

\begin{figure}
  \centering
  \includegraphics[width=1\linewidth,height=0.4\textwidth]{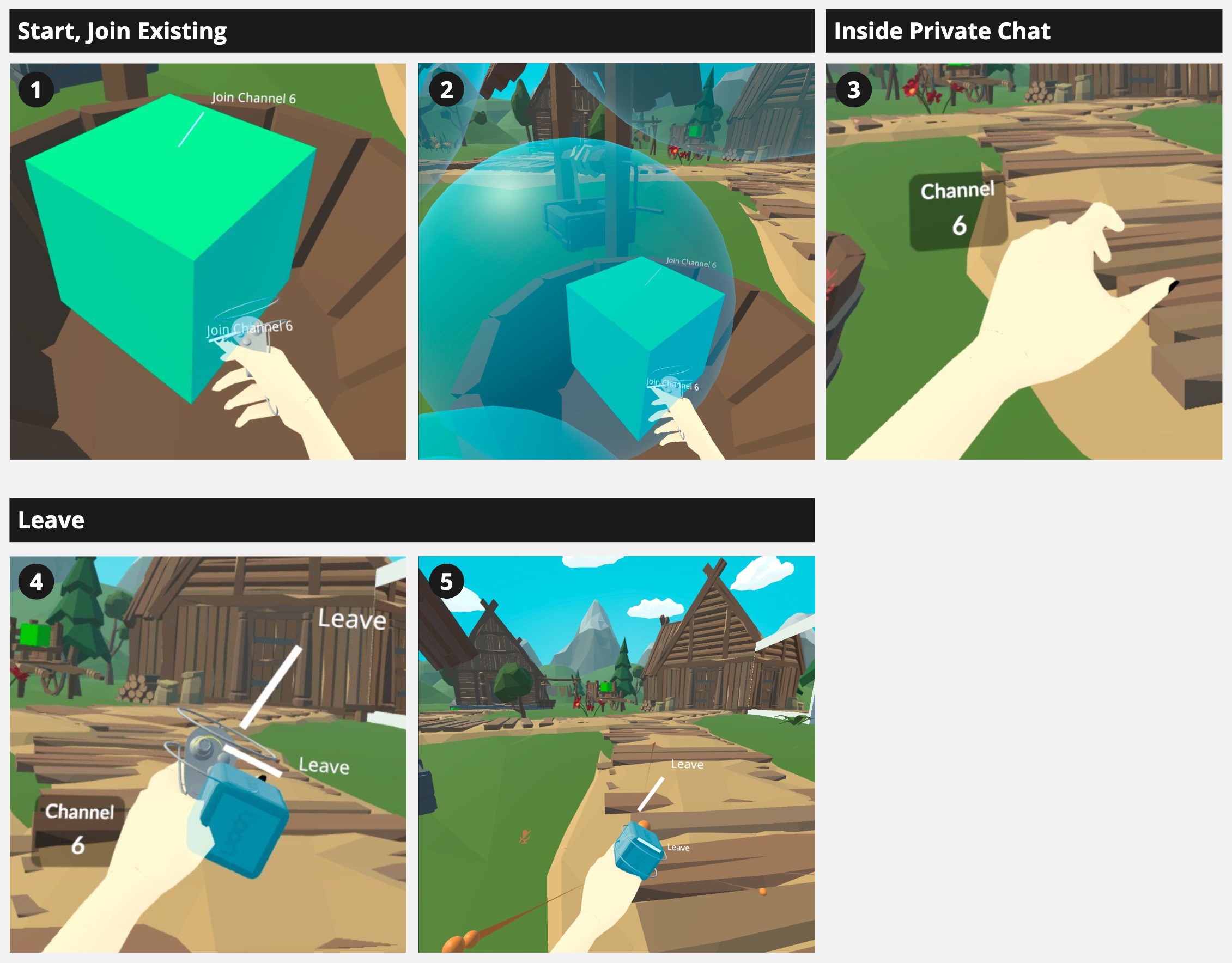}
  \caption{How to Interact with the HushHub System: 5 steps of interacting with the HushHub system. The box in step 1 is the button to join a private channel, which is later refered as the "enter button". When users suspend their controller upon the button, a label showing the channel number will appear to the user. Step 2 shows the visual effect when a user click an enter button to join a private channel. The visual effect is only visible to the user initiating the chat. Step 3 shows what it looks like within a private channel: a UI referred as the "following UI" in the main text will always appear above user's left controller, representing that the user is in a private channel. Step 4 shows the exit button and labels which will only appear when a user in private channel hold the "Grab" button in the left controller. Step 5 shows the particle visual effects when the user interact with the exit button.}
  \Description{.}
  \label{fig:interactions}
\end{figure}

\subsection{Design}

\begin{table*}
    \caption{Research Questions, Problems and Solutions}
    \label{tab:problems-solutions}
    \begin{tabular}{p{0.30\textwidth} | p{0.30\textwidth} | p{0.30\textwidth}}
        \toprule
        Concern & Specific Problem & HushHub's Solution \\
        \midrule
        Lack of Control Over Self-disclosure &  No way to limit your voice to specific people. No spontaneous whispering and private conversation allowed in current social VR & Private Voice Channels allow users to choose who can hear them. \\
        Unexpected Disruption &  In current social VR, users cannot prevent unwelcome people from entering your conversation & Outsiders are unaware of which private channel a user is in, making it difficult for unwelcome individuals to interrupt. \\
        Cannot close a private conversations in VR with convenience & With distinct audio boundaries, users might face situations where they need to stop a private chat immediately to respond to a user outside of the private channel. & Each player has a "Leave" button, accessible at all times, allowing them to exit a private channel seamlessly. \\
        Cannot follow social context outside private chat & People initiating private conversations are afraid of becoming unaware of the public context, creating difficulties in re-joining the public after private chat & Users in private channels can still hear external conversations, ensuring they remain aware of the broader context. \\
        \bottomrule
    \end{tabular}
\end{table*}

We propose some common concerns and outline how HushHub addresses these issues. Table 1 summarizes the concerns, specific problems, and the corresponding features of HushHub designed to tackle these challenges. This structured approach ensures that the features of HushHub align with user needs and address key barriers to comfortable and fluid social interactions in the Metaverse. 

We now provide a detailed explanation of how the three user actions associated with the private chat functionality address the previously identified concerns.

\subsubsection{Start Chat}
As illustrated in Fig.\ref{fig:interactions} Step 1, users can approach an Enter Button located in the virtual environment and interact with it to join the corresponding channel (e.g., pressing Button 6 adds the user to Channel 6). Upon interacting with the Enter Button (Fig.\ref{fig:interactions} Step 2), the system provides sound and visual effects exclusively visible and audible to the user indicating that the user has successfully joined the channel. For others, there is no observable change in the user’s behavior or appearance, except that they can no longer hear the user's voice. This design ensures that a user can enter a channel discreetly without interrupting ongoing discussions.

\subsubsection{Join Existing Chat}
While inside a channel, the user sees a floating panel (Fig.~\ref{fig:interactions} Step 3) that displays the channel they joined. This panel is visible only to the user who is in private chat and it follows the user as he/she moves around in the virtual world. Outsiders who wish to join the user’s channel can do so by interacting with the same Enter Button. However, once the user moves away from the Enter Button, it becomes impossible for others to determine which channel the user is in, thereby preserving their privacy.

\subsubsection{Leave Chat}
At any time, the user can hold the Grab action in the controller to reveal an Exit Button (Fig.\ref{fig:interactions} Step 4). This button is visible only to the user and disappears when the user releases the Grab action. This mechanism acts as a subtle and intuitive "confirmation screen," avoiding unintended exits that could disrupt the user’s privacy. Upon interacting with the Exit Button (Fig.\ref{fig:interactions} Step 5), the system provides feedback in the form of a sound effect and an orange bursting particle effect as indicator signals. After exiting, the user’s voice becomes audible to everyone in the environment again. This design allows users to leave the channel at their discretion, ensuring they retain full control over their interactions. By enabling users to exit voluntarily, the system minimizes the risk of misuse, such as being trapped in a conversation with an unwanted participant. 

Overall, these interactions provide an intuitive, private, and user-centric experience, supporting seamless entry, participation, and exit from private conversations.

\subsection{Implementation}

\begin{figure*}
  \centering
  \includegraphics[width=0.8\linewidth,height=0.4\textwidth]{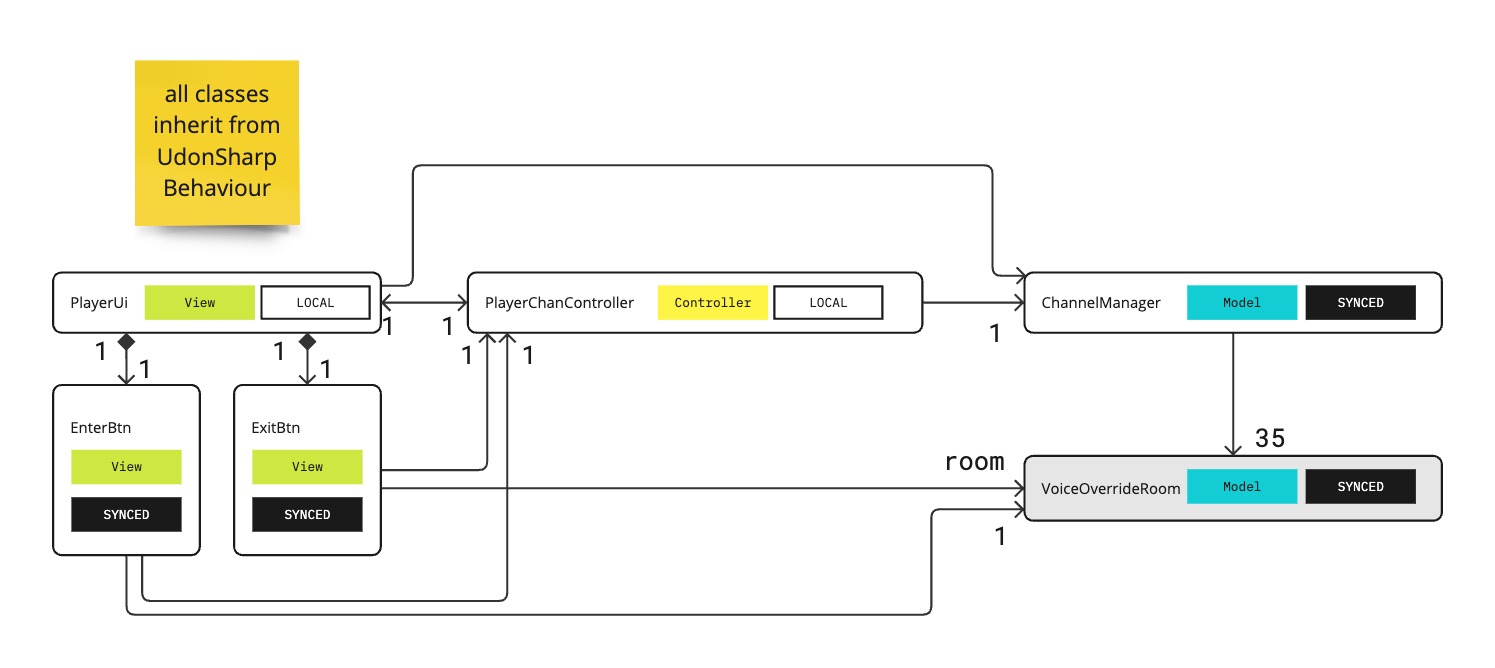}
  \caption{Class Diagram of HushHub System.}
  \Description{How classes in HushHub relate to each other with their members and methods.}
  \label{fig:classdiagram}
\end{figure*}

Rather than building a standalone VR application, we chose to integrate our project into VRChat, a well-established social VR platform. This approach allowed us to allocate more resources toward the design and development of the private chat system, bypassing the complexities associated with creating an entire application. This ensured ease of adoption and accessibility for future iterations of the project.

HushHub is implemented as a virtual world within the VRChat platform, utilizing the Unity engine. Unity employs a Mono runtime for scripting \cite{unity_mono}, and VRChat extends its functionality through the UdonSharp API \cite{merlin_2023}, which we used to develop scripts controlling gameplay features.

Similar to Unity's standard practices, UdonBehaviours—custom scripts derived from the MonoBehaviour class—are attached to GameObjects to define their behavior. The VRChat Udon API calls methods on these components, allowing interaction with user inputs, modification of properties over time, and the triggering of events across scripts. These UdonBehaviours also expose networked events that facilitate communication and interactions between players.

VRChat’s networking capabilities are based on Photon, with networked events implemented as Photon RPCs \cite{networking_2023}. For voice interactions, we utilizes the UdonVoiceUtilities package from GitHub \cite{guribo_2024}, which supports networked voice channels. Our scripts build upon and directly interact with this package to implement the specific functionalities required for HushHub’s private chat system.

The HushHub system employs the Model-View-Controller (MVC) software design pattern. This pattern segregates the responsibilities of different classes, simplifying maintenance and extensibility. The Model represents the underlying data and logic of the chat system, the View manages the visual representation of the virtual environment, and the Controller handles interactions between users and the system. (Fig.~\ref{fig:classdiagram})
\begin{itemize} 
\item {Model: \verb|ChannelManager.cs|}
\item {View: \verb|PlayerUI.cs|, \verb|EnterBtn.cs| and \verb|ExitBtn.cs|}
\item {Controller: \verb|PlayerChannelController.cs|}
\end{itemize}

The visual environment of HushHub is designed with a low-poly aesthetic, sourced from the Unity Asset Store \cite{gigel_2019}. This design is both visually appealing and efficient for rendering, enhancing user experience while minimizing performance costs. To further optimize performance, invisible box colliders were used for the houses and floors instead of resource-intensive mesh colliders.

The virtual environment includes evenly distributed Enter Channel Buttons, allowing ample options for participants to join private channels. For testing purposes, seven channels were made available, with each session accommodating up to ten participants. The virtual world and its code were uploaded to VRChat servers, and private instances were launched during testing sessions.

\section{Research Methodology}

\subsection{Participants}
A user study was conducted with Institutional Review Board approval, using convenience sampling. The 21 participants, aged 18 to 30, were recruited from a university VR Interaction Design class and personal networks, with prior experience in online gaming or social media required. The sample comprised 71.4\% males and 28.6\% females, evenly distributed across Singapore, Malaysia, and China. Only two participants had prior VRChat experience. Participants were randomly divided into three groups (6, 6, and 8 members) to test the Hushhub system. Some participants knew each other beforehand (by coincidence, each group has two people who knows each other in advance), while others met for the first time during the pre-trial briefing.

\subsection{Pre-Trial Survey}

A pre-trial survey was sent out to participants for several key functions. First, it obtained informed consent from participants and explained our data usage protocols to ensure their privacy. Second, participants indicated their availability for the user study and whether they owned a VR headset. This information allowed us to create a randomized schedule for testing sessions. For those without personal VR headsets, we arranged for them to participate at a pre-booked on-campus venue during their scheduled time slots.

\subsection{Trials}
The user study spanned three days, each day with one trial group. Participants with headsets joined a designated VRChat instance from their own locations at a pre-arranged time, coordinated via Zoom. Participants without headsets attended the study at a university’s venues, where they were provided with Meta Quest 2 headsets and assigned to seperated rooms during the trial.

A 15-minute pre-trial briefing was conducted to explain the study’s objectives and provide practical guidance. The topics included the purpose of the study, how to use VR headsets and controllers, navigation in VRChat, joining the designated VRChat world, and using the private conversation system. Participants were asked to explore VRChat and the Hushhub system freely without specific indications. 

The trial session lasted 30 minutes, during which participants engaged in conversations and explored the VRChat world (eg, choosing avatars). Participants were initially in the public channel and were provided with the option to utilize the private conversation system as the trial starts. During the trial, all VRChat common features are available as the default features in VRChat instances all remained unchanged in the trial world instance. Following the trial, participants completed an post-trial survey, which required approximately 15 minutes.

\subsection{Post-Trial Survey}

The post-trial survey assessed both the quantity and quality of the interactions participants had during the study. They were first asked if they had conversations with others and whether they tested the private chat feature during the conversation, which all participants replied yes. We also collect user feedback that answers our \textbf{RQ}s.

\section{Results}

\begin{figure}
  \centering
  \includegraphics[width=0.8\linewidth,height=0.4\textwidth]{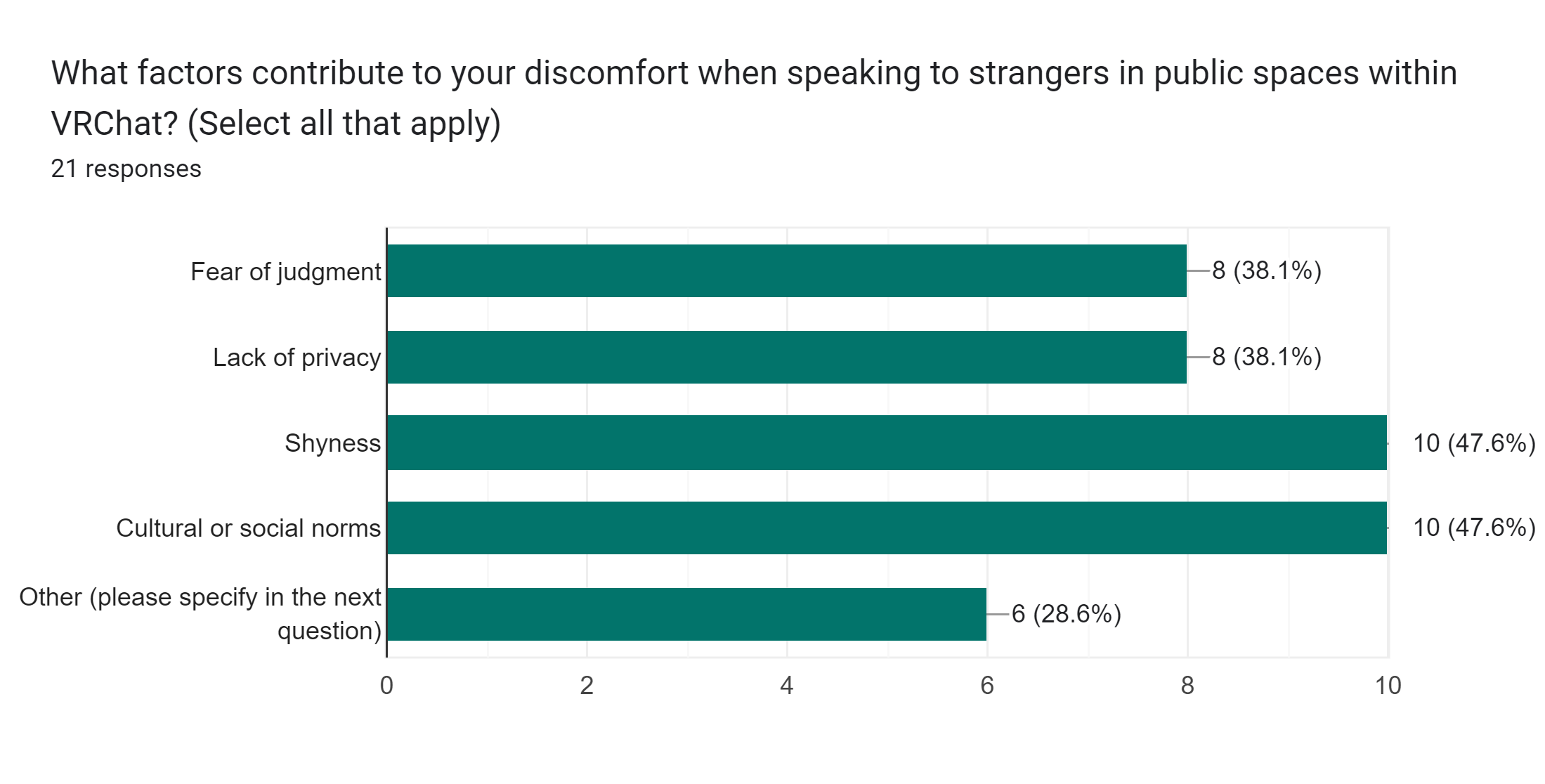}
  \caption{Factors contributing to participants' discomfort when speaking to strangers in public in VRChat.}
  \Description{.}
  \label{fig:factors}
\end{figure}

\begin{figure}
  \centering
  \includegraphics[width=1.0\linewidth,height=0.4\textwidth]{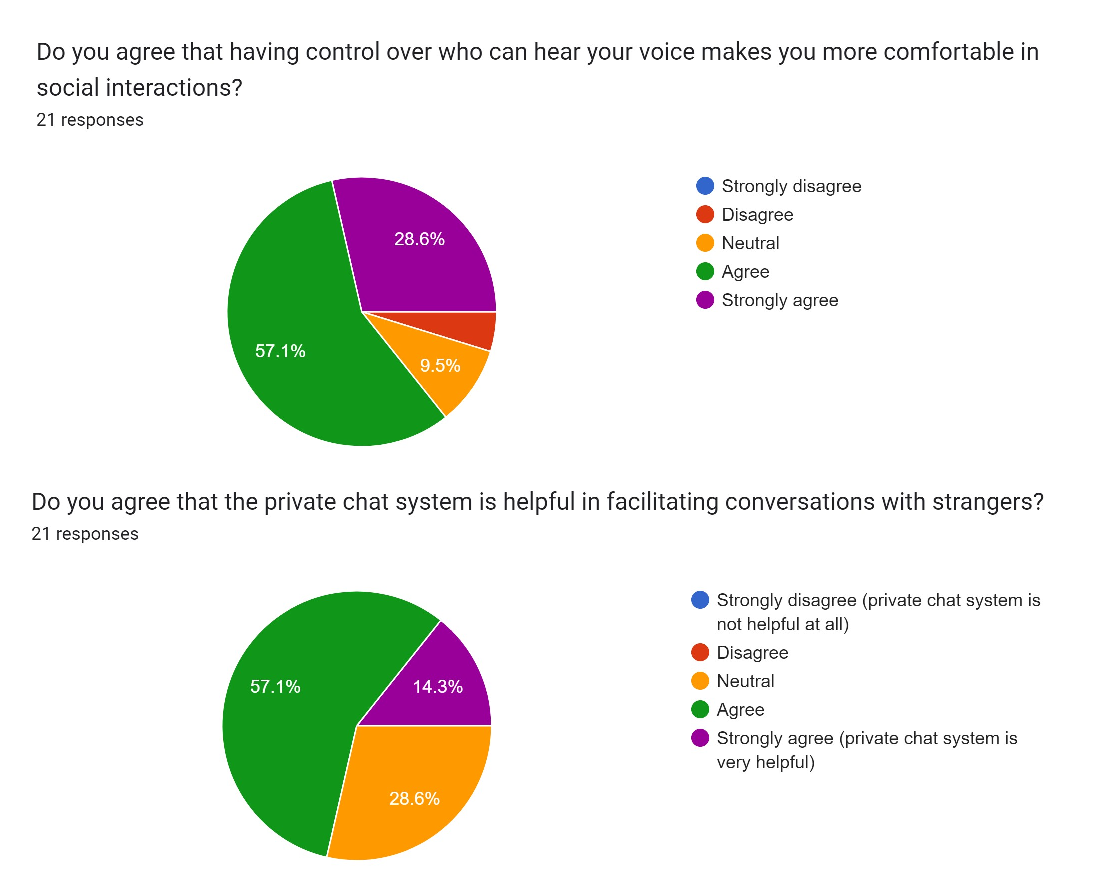}
  \caption{Review on having control over who can hear the participants (upper graph) and whether private conversation system is helpful (lower graph).}
  \Description{.}
  \label{fig:ishelpful}
\end{figure}

\begin{figure}
  \centering
  \includegraphics[width=0.6\linewidth,height=0.4\textwidth]{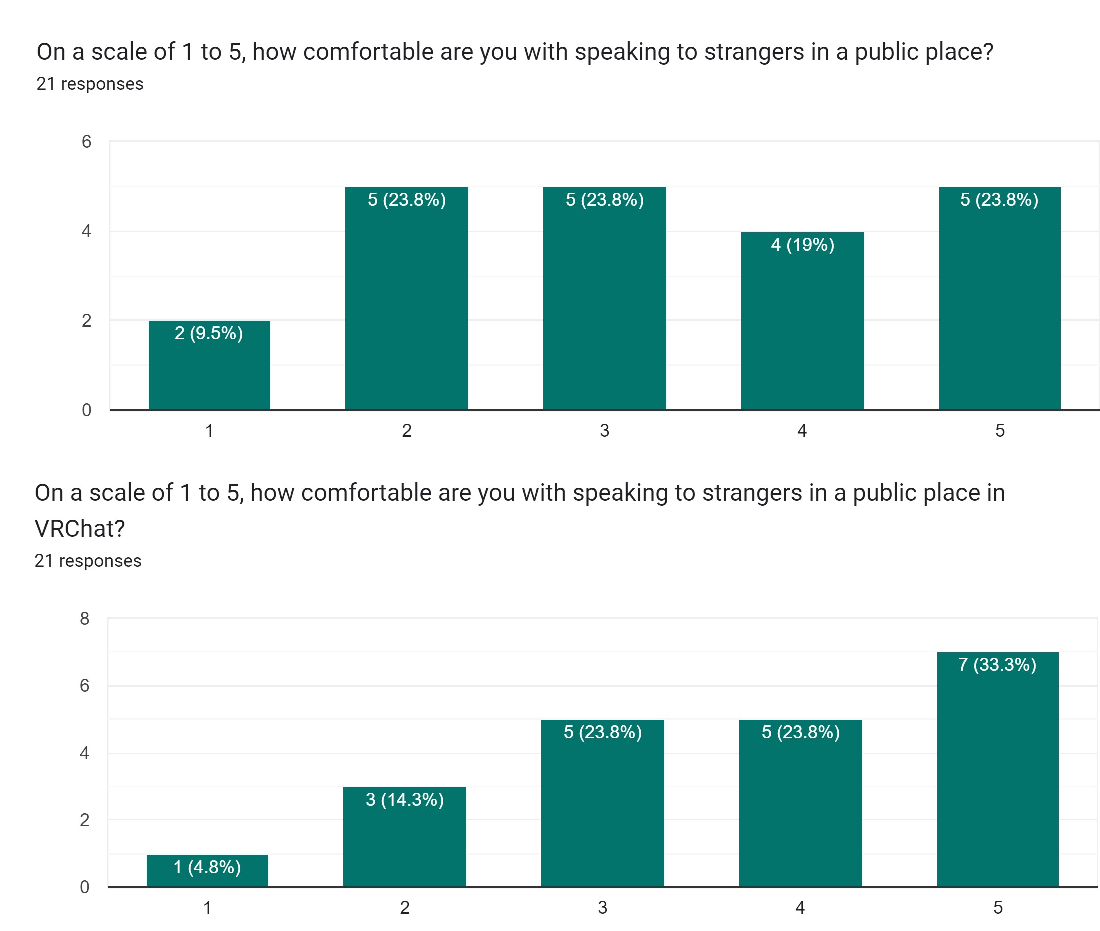}
  \caption{Comfort Level on talking to strangers in two scenarios, 1 being very uncomfortable, 5 being very comfortable. The upper graph shows the comfort level when speaking to strangers in public in reality with Median = 3 and IQR = 2.5; the lower grpha shows the comfort level when speaking to strangers in VRChat public space with Median = 4 and IQR = 2. A paired t-test ($\alpha$ = 0.05) was conducted using the two graphs}
  \Description{.}
  \label{fig:comfortlevel}
\end{figure}

\subsection{Survey Data}

The results of the post-trial survey provide insights into our research inquiries. Data were collected on participants' perceptions regarding the challenges when interacting with others in virtual reality (VR) environments (Fig.~\ref{fig:factors}). The analysis showed an even distribution of responses across several categories, including "Fear of judgment" (8/21), "Lack of privacy" (8/21), "Shyness" (10/21), and "Cultural or social norms" (10/21). Specific reasons cited by participants included the inability to interpret certain social cues ("\textsl{Unable to see certain things about the person (their facial expression, etc.)}," noted Participant 4 (P4)), feelings of exclusion ("\textsl{Too many kids / other cliques which makes me feel like an outsider}," P6), and concerns about identity exposure ("\textsl{... I can be a skeleton in VR but once I talk all the anonymity is gone}," P18). This provides answers to \textbf{RQ1} about the challenges users face in public interactions. Among the answers, lack of privacy features is a highly concerned factor besides user characteristics, proving that the deficiency in current social VR platforms degrades communication effectiveness.

The effectiveness of the private conversation system was also assessed (Fig.~\ref{fig:ishelpful}). The majority of participants (18/21) agreed that having control over who can hear their voice enhanced their comfort in social interactions. Additionally, most participants (15/21) believed that the system increased their motivation to communicate with strangers. These results suggest that the ability to initiate side conversations—without leaving the shared environment—can lower social friction and encourage spontaneous interaction, especially among unfamiliar users. Further qualitative insights were revealed in open-ended responses, where participants highlighted several key benefits of the private chat feature. For instance, participants felt less targeted ("\textsl{outside the bubbles sometimes I felt like I was just talking to the void}," P19), experienced a more personal connection ("\textsl{Outside is less personal, more group oriented}," P1), and appreciated the ability to maintain focus during private conversations ("\textsl{I feel that talking inside a private chat can let me focus on the person that I am talking to}," P9). These findings indicate that the private conversation system positively impacts users' comfort, motivation, and quality of interactions in VRChat environments, answering \textbf{RQ2}.

\subsection{Additional Feedback}

This section presents insights derived from survey questions that provide valuable context to our primary research focus.

\textsl{5.2.1 Comparing Users' Comfort Levels of Public Interactions in Reality and VR settings}. A paired t-test ($\alpha$ = 0.05) was conducted to compare participants' comfort levels when engaging with strangers in public in real-life scenarios versus within VRChat (Fig.~\ref{fig:comfortlevel}). The results revealed a significant difference, indicating that participants felt more comfortable speaking publicly to strangers in VRChat (Median = 4, IQR = 2) compared to real-life settings (Median = 3, IQR = 2.5, p < 0.05). Participants attributed this increased comfort to the anonymity provided by VR ("\textsl{Inside u don’t see their eyes which means less awkward}," P3) and the perceived friendliness of others' avatars ("\textsl{I just feel people appear nicer to me when they are in their avatars}," P17).

\textsl{5.2.2 Participant Suggestions for Improving the Private Conversation System}. Participants provided constructive feedback on areas for improvement within the private conversation system. Some noted challenges related to inadvertently remaining in private channels due to the subtle design of the channel interface. For instance, one participant remarked, "\textsl{More bulky avatars end up hiding the channel UI, consider moving it to the top of the view ...}," (P13). Others highlighted difficulties in identifying other users within the same private conversation channel, suggesting the addition of features to enhance clarity. As one participant suggested, "\textsl{Maybe a way of knowing who else is in the channel}," (P21).

These findings highlight both the perceived value of VR platforms and private chat systems in social contexts, as well as specific areas for refinement to enhance user experience and functionality.

\section{Conclusion}

The study addresses the proposed \textbf{RQ}s by demonstrating how a private communication system can bridge a key gap in current social VR platforms—the lack of support for private conversations that naturally occur in real-world group settings. We emphasizes the importance of replicating authentic social dynamics through private and parallel communication channels in the Metaverse. Building on existing mechanisms of mainstream social VR platform, our implemented system, Hushhub, allows users to initiate private conversations within shared VR spaces without disrupting the ongoing group discourse. The evaluation of HushHub highlights its effectiveness in enhancing the realism, flexibility, and depth of social interactions in virtual environments, offering a practical step toward making social VR experiences more closely aligned with real-world social behaviors.

Future research will aim to explore user behavior surrounding self-disclosure in virtual reality environments in greater depth, specifically planning for more diverse participant samples, longitudinal studies, and large scale studies. This will enable a more nuanced understanding of social dynamics in virtual spaces. This evolving knowledge base will be critical for designing even more inclusive and engaging VR experiences. HushHub represents a significant step forward in creating a more equitable and enjoyable Metaverse, paving the way for enhanced privacy, inclusivity, and meaningful social interactions for all users.

\bibliographystyle{ACM-Reference-Format}
\bibliography{xwei-jerry-ian}

\end{document}